\documentclass[twocolumn,amsmath,amssymb,showpacs]{revtex4}
\usepackage{dcolumn}
\usepackage{color}
\usepackage{amsmath}

\usepackage[dvips]{graphicx}

\begin{document}

\begin{sloppy}

\title{Truncated-Bloch-wave solitons in optical lattices}

\author{Jiandong Wang$^{1,2}$, Jianke Yang$^2$, Tristram J. Alexander$^3$, and Yuri S. Kivshar$^3$}

\affiliation{$^1$ College of Physical Electronics, University of Electronic Science and Technology of China, Chengdu 610054, China \\
$^2$ Department of Mathematics and Statistics, University of Vermont, Burlington, VT 05401, USA\\
$^3$Nonlinear Physics Center, Research School of Physical Sciences
and Engineering, Australian National University, Canberra ACT 0200,
Australia}

\begin{abstract}
We study self-trapped localized nonlinear states in the form of
truncated Bloch waves in one-dimensional optical lattices, which
appear in the gaps of the linear bandgap spectrum. We demonstrate
the existence of families of such localized states which differ by
the number of intensity peaks. These families do not bifurcate from
the band edge, and their power curves exhibit double branches.
Linear stability analysis demonstrates that in deep lattice
potentials the states corresponding to the lower branches are
stable, whereas those corresponding to the upper branches are
unstable, independently of the number of peaks.
\end{abstract}

\pacs{03.75.Lm, 05.45.Yv}
% 03.75.Lm: Tunneling, Josephson effect, Bose-Einstein condensates in periodic potentials,
%           solitons, vortices and topological excitations
% 05.45.Yv: Solitons

\maketitle

The theoretical study of localized states in periodic potentials is
driven by many experimental demonstrations in nonlinear
optics~\cite{optics} and the physics of Bose-Einstein condensates
(BECs)~\cite{BEC_new,BEC1,BEC3,BEC4}. One of the most striking
effects is the possibility of spatial localization due to an
interplay between periodicity and nonlinearity, even when the
nonlinearity is repulsive. The corresponding localized states have
long been known in nonlinear optics as {\em gap
solitons}~\cite{book} whose frequencies lie inside the gaps of the
linear band-gap spectrum of a periodic structure. Similar objects
have been observed in atomic physics, where the experimental
generation of BEC gap solitons proved to be a challenge due to the
requirement of low atom numbers and densities~\cite{BEC4}. The
physics of such localized states is well understood, and usually the
gap solitons appear through bifurcations of the localized states
from the edges of the bands of periodic solutions~\cite{gap_theory}.

Seemingly different localized states with steep edges and a large
number of atoms populated evenly in a number of adjacent lattice
sites was observed a few years ago in BEC loaded into an optical
lattice~\cite{BEC_gap}. It was observed that in a deep
one-dimensional optical lattice the BEC wave packet does not
diffuse,  but instead its initial expansion stops and the width
remains finite. This effect has been attributed to the self-trapping
mechanism of energy localization and the existence of novel types of
broad nonlinear localized states, in the form of truncated nonlinear
Bloch waves in the gaps of the matter-wave linear
spectrum~\cite{our_prl}. In particular, Alexander~{\em et
al.}~\cite{our_prl} revealed that these robust nonlinear localized
states may exist in all dimensions and can have arbitrary extension
within the lattice. They seem to provide an important missing link
between gap solitons and nonlinear Bloch waves, and can be termed
{\em truncated-Bloch-wave solitons}.

In this paper, we consider a model for the Bose-Einstein condensate
in a one-dimensional optical lattice under repulsive nonlinearity,
and systematically study the existence and stability of the
truncated-Bloch-wave solitons. On the question of existence, we
reveal that these solitons exist as countable families which are
characterized by the number of intensity peaks. These solution
families do not bifurcate from the edges of Bloch bands, and their
power curves exhibit double branches. If the depth of the lattice
potential falls below a certain threshold value, these
truncated-Bloch-wave solitons cease to exist. We also demonstrate
that in a deep lattice potential these solitons on the lower
branches of the power curves are stable, while those on the upper
branches of the power curves are unstable, independent of the number
of intensity peaks. In a shallow lattice potential, both branches of
these solitons are linearly unstable. These results provide a solid
background for the earlier experimental observations of broad
localized states in BECs~\cite{BEC_gap} as well as the corresponding
numerical studies~\cite{our_prl}, and they reveal the specific role
these novel types of self-trapped localized states play as a link
between nonlinear Bloch waves and gap solitons in periodic
potentials.

The physical system we consider is that of a Bose-Einstein
condensate loaded into a one-dimensional optical lattice under
repulsive nonlinearity~\cite{BEC4,BEC_gap}. The mathematical model
for this system is the one-dimensional Gross-Pitaevskii equation
with a lattice potential \cite{BEC_model}
\begin{equation} \label{U}
iU_t+U_{xx}-(V_0\sin^2x)U-|U|^2U=0,
\end{equation}
where $U$ is the mean-field wavefunction of the BEC, and the lattice
potential is $\pi$-periodic with potential depth $V_0$. We have used
the same normalization as that found in Ref.~\cite{our_prl} to make
the model dimensionless, with all quantities in units of
characteristic scales of the lattice.  In particular $x$ and $t$ are
in units of $a_L = d/\pi$ and $\omega_L^{-1} = \hbar/E_{rec}$
respectively, where $d$ is the period of the lattice and $E_{rec} =
\hbar^2/2ma_L^2$ is the recoil energy for an atom of mass $m$
absorbing a lattice photon.  The depth of the lattice, $V_0$, is in
units of $E_{rec}$.
%The physical density in atoms.cm$^-3$ may be obtained by multiplying the normalized density $|U|^2$ by $2\times10^{-6}/8\pi a_L^2|a_s|$ where $a_s$ is the s-wave scattering length.
We have assumed that the condensate is tightly confined in the transverse dimensions to use the standard dimensionality reduction procedure to reduce the dynamics to an effectively one-dimensional model.
%To obtain the total number of particles from the normalized one-dimensional particle number $N=\int_{-\infty}^\infty|u|^2dx$ defined relative to the stationary profile $u(x)$ (discussed below) one must multiply $N$ by $\hbar/(4|a_s|m\omega_\perp a_L)$ where $\omega_\perp$ is the transverse trap frequency.

Solitons in
this system are sought in the form of $U(x, t)=u(x)e^{-i\mu t}$,
where $\mu$ is the chemical potential, and the spatial function
$u(x)$ is real-valued and satisfies the reduced equation
\begin{equation} \label{u}
u_{xx}-(V_0\sin^2x) u-u^3=-\mu u.
\end{equation}
In this paper, we first consider a deep lattice potential with
$V_0=6$, where the existence and linear stability of
truncated-Bloch-wave solitons will be determined in detail. These
solitons and their stability in shallower potentials will be
addressed toward the end of the paper.

To understand the origin of the truncated-Bloch-wave solitons, we
first examine periodic nonlinear Bloch waves which originate from
linear Bloch waves at the band edges. In a deep potential with
$V_0=6$, the linear dispersion curves of Eq. (\ref{u}) can be found
in Fig.~\ref{fig1}(a) of Ref.~\cite{ShiYang}. At the lower edge of
the first Bloch band [which is 2.0632 in Fig. \ref{fig0}(a)], Eq.
(\ref{u}) admits an infinitesimal (linear) Bloch wave which is
$\pi$-periodic, and its adjacent intensity peaks are in-phase with
each other. This linear Bloch wave (in arbitrary units) is shown in
Fig.~2(1) of Ref.~\cite{ShiYang} and appears similar to the profile
shown in Fig. \ref{fig0}(b) here. When $\mu$ increases from this
band edge (into the first Bloch band), this linear Bloch wave
bifurcates into a nonlinear $\pi$-periodic Bloch wave with a finite
amplitude (under the present repulsive nonlinearity). Numerically we
can easily compute these nonlinear Bloch waves, and their amplitudes
versus the chemical potential $\mu$ are displayed in
Fig.~\ref{fig0}(a). Here the amplitude of a nonlinear Bloch wave
$u(x)$ is defined as the maximum value of $u(x)$. At two
representative chemical potentials $\mu=2.16$ and $4$ (the former
lies inside the first Bloch band, and the latter lies inside the
first bandgap), the corresponding nonlinear Bloch waves are
displayed in Figs.~\ref{fig0}(b,c), respectively. As can be seen, as
$\mu$ increases from the lower band edge, the amplitude of the
nonlinear Bloch wave also increases.

\begin{figure}[t]
\begin{center}
\includegraphics[width=8.5cm]{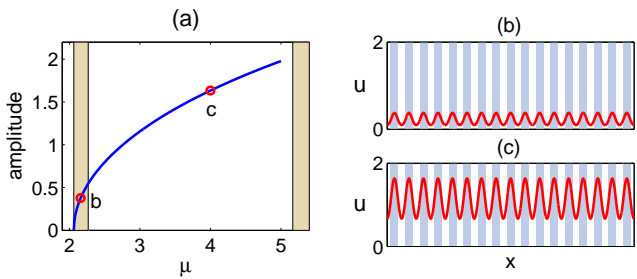}
\end{center}
\caption{Nonlinear Bloch waves bifurcated from the lower edge of the
first Bloch band under repulsive nonlinearity ($V_0=6$). (a) The
amplitude versus the chemical potential diagram; (b, c) nonlinear
Bloch waves at two chemical potentials marked in (a). The shaded
stripes represent low-potential regions.} \label{fig0}
\end{figure}

Now we truncate these infinitely extended nonlinear Bloch waves to a
finite number of intensity peaks (with the $\mu$ value fixed), and
ask whether the truncated Bloch waves can develop exponentially
decaying tails at large $|x|$ values so that they become localized
soliton solutions of Eq. (\ref{u}). When the chemical potential lies
inside the first Bloch band [see Fig. \ref{fig0}(b)], the truncated
Bloch waves will still possess continuous-wave tails and are thus
still infinitely extended. But when the chemical potential lies inside
the first band gap [see Fig. \ref{fig0}(c)], then due to repeated
Bragg reflections, the truncated Bloch waves \emph{can} decay
exponentially at large $|x|$ values and thus become localized
solitons. To confirm this, we numerically compute soliton solutions
of Eq. (\ref{u}) by the squared-operator iteration methods developed
in Ref.~\cite{SOM} using the truncated nonlinear Bloch waves as initial
conditions. When truncating the Bloch wave of Fig.~\ref{fig0}(c) to
three and seven peaks, we indeed obtained soliton solutions which
are displayed in Fig.~\ref{fig1}(b,d) respectively. These solitons
closely resemble the corresponding nonlinear Bloch wave [see Fig.
\ref{fig1}(d) for comparison], but their tails are now exponentially
decaying rather than infinitely extended. If the Bloch wave of Fig.
\ref{fig0}(c) is truncated to any other number of peaks, the
corresponding soliton state could be obtained as well. Thus a
countable number of solitons can be found from different truncations
of a nonlinear Bloch wave (at the same $\mu$ value). These solitons
are the ones which we termed \emph{truncated-Bloch-wave solitons}
above. They were first observed in BEC experiments~\cite{BEC_gap}
and then found numerically in Ref.~\cite{our_prl}.

\begin{figure}[t]
\begin{center}
\includegraphics[width=8.5cm]{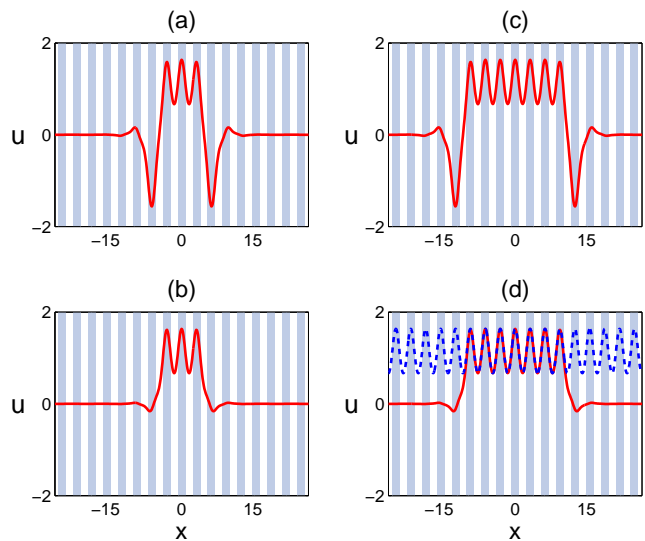}
\end{center}
\caption{Truncated-Bloch-wave solitons on the upper and lower power
branches of the three-peak (left column) and seven-peak (right
column) families in the first bandgap ($V_0=6$). The chemical
potentials of these solitons are marked in Fig.~\ref{fig2}. In (d),
the corresponding nonlinear Bloch wave (dashed curve) is also shown
for comparison [see also Fig. \ref{fig0}(c)]. } \label{fig1}
\end{figure}

When the chemical potential changes, the truncated-Bloch-wave
solitons will deform. Hence each truncated-Bloch-wave soliton
generates a continuous family of such solitons parameterized by the
chemical potential. We will call the solution family of the
three-peak truncated-Bloch-wave soliton in Fig. \ref{fig1}(b) the
three-peak family, and that of the seven-peak soliton in Fig.
\ref{fig1}(d) the seven-peak family, and so on. Important questions
which have not been addressed before are: what solutions are
contained in each family? and what do their power diagrams look
like? Here the power of a soliton is defined as
$P=\int_{-\infty}^\infty |u|^2dx$ as usual. To address these
questions, we have numerically obtained the entire families of
two-peak to seven-peak solitons. The power diagrams of three-peak
and seven-peak families are displayed in Fig. \ref{fig2} (power
curves for other-peak families are similar). These power diagrams
show three important common features about truncated-Bloch-wave
solitons. The first feature is that these solitons exist only inside
the bandgap, but not inside a Bloch band. This contrasts with the
associated nonlinear Bloch waves which exist in both the bandgap and
the Bloch band [see Fig. \ref{fig0}(a)]. The second feature is that
these solitons do not bifurcate from the (upper) edge of the (first)
Bloch band, thus they are fundamentally different from the gap
solitons studied before \cite{gap_theory} which do bifurcate from
edges of Bloch bands. The reason these solitons can not bifurcate
from the upper edge of the first Bloch band is obviously that
adjacent peaks in these solitons are in-phase, which does not match
the Bloch wave at the band edge (where adjacent peaks are
out-of-phase). Note that these solitons are closely related to
nonlinear Bloch waves which bifurcate from the lower edge of the
first Bloch band (see Fig. \ref{fig0}). But unlike those nonlinear
Bloch waves, these solitons can not exist inside the first Bloch
band, thus they can not bifurcate from the lower edge of the first
Bloch band.  The third common feature of the truncated Bloch-wave
solitons is that their power diagrams exhibit double branches. This
double-branch phenomenon has been reported for dipole and vortex
solitons in periodic potentials recently \cite{jianke_pra,torner}.
Now we know that this phenomenon will always occur for any soliton
family which does not bifurcate from edges of Bloch bands (or edges
of the continuous spectrum) as is the case for truncated-Bloch-wave
solitons. The three-peak and seven-peak solitons shown in
Fig.~\ref{fig1}(b,d) lie on the lower branches of their solution
families (more specifically at points `b, d' marked in Fig.
\ref{fig2}). On the upper branches, the solitons will develop two
extra intensity peaks at the two edges of the solitons, and these
extra peaks are out-of-phase with the other intensity peaks in the
interior. Two examples on the upper branches of the three-peak and
seven-peak families are displayed in Fig. \ref{fig1}(a,c), and their
positions on the power diagrams are marked in Fig.~\ref{fig2}. It is
noted that these solution families of various peaks are distinct
from each other, with no s-shaped bifurcations connecting them near
the band edge in the power diagram. When the chemical potential
$\mu$ approaches the right end of the first bandgap, all solutions
become less localized. If $\mu$ crosses this right end into the
(second) Bloch band, solutions become delocalized due to resonance
with the Bloch modes, and the powers of the solutions become
infinite.

\begin{figure}[t]
\begin{center}
\includegraphics[width=8.5cm]{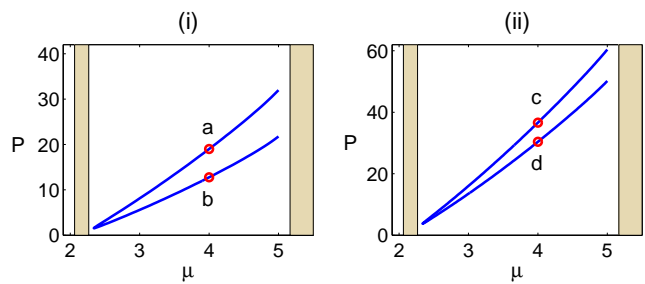}
\end{center}
\caption{Power curves of the three-peak (a) and seven-peak (b)
families of truncated-Bloch-wave solitons ($V_0=6$). The solutions
at marked points are displayed in Fig. \ref{fig1}. } \label{fig2}
\end{figure}

Stability of these truncated-Bloch-wave solitons is clearly an
important issue, and this issue has not been considered before in
the literature. Here we investigate the linear stability of these
solitons. Following the standard procedure, we perturb the soliton
as $U(x, t)=e^{-i\mu t}\left[u(x)+(v(x)-w(x))e^{\lambda
t}+(v^*(x)+w^*(x))e^{\lambda^* t}\right]$, where $v, w \ll 1$ are
normal-mode perturbations, $\lambda$ is the eigenvalue of the normal
mode, and the superscript `*' represents complex conjugation.
Inserting this perturbed soliton into the evolution equation
(\ref{U}) and dropping higher-order terms in $(v, w)$, a linear
eigenvalue problem $L(v, w)^T=\lambda (v, w)^T$ is obtained. Here
$L$ is the linearization operator, and the superscript `T'
represents the transpose of a vector. Linear stability of a soliton
is determined by the spectrum of the linearization operator $L$, and
the existence of any eigenvalue with a positive real part implies
linear instability of the soliton. We have computed this spectrum
for truncated-Bloch-wave solitons of various families in the deep
potential of $V_0=6$ (see Figs.~\ref{fig1} and \ref{fig2}) by the
Fourier collocation method \cite{Boyd}. We have found that for all
these solution families, solitons on the lower branches are always
stable, and those on the upper branches always unstable. To
illustrate, we pick the two solitons in Fig.~\ref{fig1}(c, d), which
lie on the upper and lower branches of the seven-peak family in
Fig.~\ref{fig2}(b). The linear-stability spectra of these two
solitons are shown in Fig.~\ref{fig3}(b,a), respectively. We see
that eigenvalues of the lower-branch soliton all lie on the
imaginary axis, indicating that this soliton is linearly stable. But
the spectrum for the upper-branch soliton contains real positive
eigenvalues, indicating its linear instability. Intuitively, the
linear instability of this upper-branch soliton is easy to
understand. Indeed, this soliton has two intensity peaks on the two
edges which are out-of-phase with the interior peaks. Thus the
structures near the edges of this soliton can be viewed as
out-of-phase dipoles. It is well known that in lattices under
attractive nonlinearity, in-phase dipoles are linearly unstable
(see, e.g., Ref.~\cite{Yangdipole}). If the nonlinearity is
repulsive as it is in the model (\ref{U}), dipole stability is
switched, and out-of-phase dipoles become linearly unstable (see
e.g. Ref.~\cite{panos}). Consequently, this upper-branch soliton in
Fig.~\ref{fig1}(c) should be linearly unstable. Regarding the linear
stability of the lower-branch soliton, however, it is less obvious
intuitively, and this result is one of the main findings of this
paper.

\begin{figure}[t]
\begin{center}
\includegraphics[width=8.5cm]{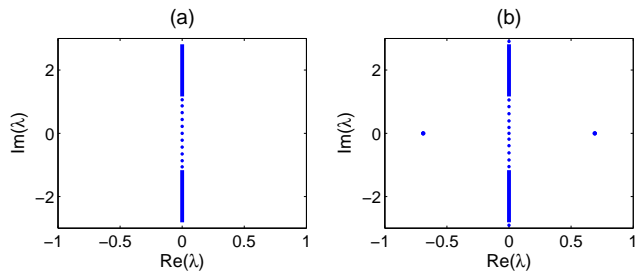}
\end{center}
\caption{Linear-stability analysis of the seven-peak family of
truncated-Bloch-wave solitons in Figs. 1 and 2 ($V_0=6$): (a)
stability spectrum of the lower-branch soliton shown in Fig.
\ref{fig1}(d); (b) stability spectrum of the upper-branch soliton
shown in Fig. \ref{fig1}(c). } \label{fig3}
\end{figure}

The above linear-stability result of lower-branch
truncated-Bloch-wave solitons in a deep lattice potential shows that
these solitons should also be nonlinearly stable under weak
perturbations. To confirm this, we take the seven-peak soliton of
Fig.~\ref{fig1}(d), and modulate its seven peak amplitudes by 10\%
random perturbations. Specifically we add to this soliton a
perturbation which is a superposition of seven Gaussian humps of the
shape $\mbox{exp}(-x^2)$ centered at the seven lattice sites of the
soliton. The amplitudes of these Gaussian humps are taken randomly
at the level of about 10\% of the soliton's amplitude. For the
realization of amplitude values $(0, -0.08, 0.06, -0.06, 0.01, 0.05,
-0.11)$ for these Gaussian humps, the nonlinear evolution of the
perturbed soliton after 60 time-unit evolution is displayed in
Fig.~\ref{fig4}. We see that the evolution is stable as expected,
showing the robustness of these new types of soliton structures in
lattice potentials. Evolution with other realizations of the
perturbations is qualitatively similar.

\begin{figure}[t]
\begin{center}
\includegraphics[width=8.5cm]{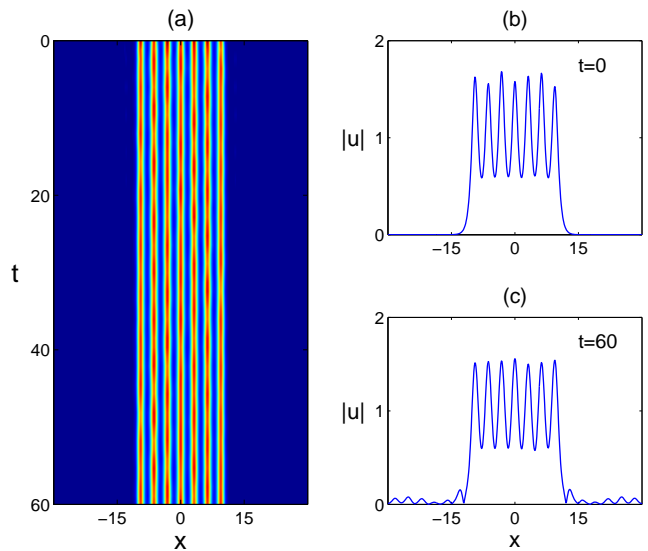}
\end{center}
\caption{Nonlinear evolution of the seven-peak truncated-Bloch-wave
soliton of Fig. \ref{fig1}(d) under 10\% perturbations: (a)
intensity ($|u|^2$) contours in the $(x, t)$ plane; (b) initial
perturbed state of $|u|$; (c) final state of $|u|$ after 60
time-unit evolution. } \label{fig4}
\end{figure}

These theoretical results appear very similar to the experimental
observation of nonlinear self-trapping of Bose-condensed $^{87}$Rb
atoms in a one-dimensional deep periodic potential~\cite{BEC_gap}.
In the experiment of Ref.~\cite{BEC_gap}, it was observed that with
a higher number of atoms, the repulsive atom-atom interaction led to
the formation of a self-trapped broad localized state with steep
edges and roughly uniform interior, closely resembling the
truncated-Bloch-wave soliton shown in Fig.~\ref{fig1}(d).  We note
however that the lattice depth used in the experiment corresponds to
$V_0 \sim 11$, much deeper than the $V_0 = 6$ considered in
Fig.~\ref{fig1}.  Furthermore, as noted in Ref.~\cite{BEC_gap}, the
experiment was not carried out in the one-dimensional regime, with
higher-order transverse effects playing an important role.  To
directly compare our results with those of experiment we may obtain
the physical density and atom number from our normalized model by
multiplying the normalized density $|U|^2$ by $2\times10^{-6}/8\pi
a_L^2|a_s|$, where $a_s$ is the s-wave scattering length, to obtain
the density in atoms.cm$^{-3}$ and multiply the power $P$ by
$\hbar/(4|a_s|m\omega_\perp a_L)$, where $\omega_\perp$ is the
transverse trap frequency, to obtain the total number of atoms.
Inherent in our one-dimensional approximation is the assumption that
the nonlinear energy at the peak density is much less than the
transverse trap energy, i.e. that $|u_{peak}|^2\hbar/(m a_L^2
\omega_{\perp}) << 1$, and as such we consider the experimental
parameters $d = 3.5\mu \mbox{m}$ and $\omega_\perp = 320 \mbox{Hz}$.
For a $^{87}$Rb BEC the seven-peak solution shown in Fig.
~\ref{fig1}(d) then contains approximately 400 atoms, with $\sim 60$
atoms per lattice period and a peak density of $\sim 2.8\times
10^{13}$atoms.cm$^{-3}$, all experimentally reasonable values.

The above theoretical results pertain to truncated-Bloch-wave
solitons in a deep lattice potential. What will happen to these
solitons if the lattice potential becomes shallow? To address this
question, we take a shallow potential with $V_0=1$, and investigate
the existence and stability properties of the truncated-Bloch-wave solitons.
For the seven-peak family, the results are summarized in
Fig.~\ref{fig5} (results for other-peak families are qualitatively
similar). In this shallow potential, we see from Fig.~\ref{fig5}(a)
that these solitons exist in a smaller region of the first gap. From
Fig.~\ref{fig5}(b,c) we see that the edges of these solitons become
less steep, their tails become longer, and solution profiles on the
upper and lower branches become more similar. From Fig.
\ref{fig5}(d), we see that the soliton on the lower branch now
becomes linearly unstable due to oscillatory instabilities caused by
complex unstable eigenvalues (this holds for other solitons on the
lower branch as well). The solitons on the upper branch are
certainly also unstable. Thus in a shallow potential, both branches
of truncated-Bloch-wave solitons are linearly unstable, which
contrasts with the deep-potential case.

\begin{figure}[t]
\begin{center}
\includegraphics[width=8.5cm]{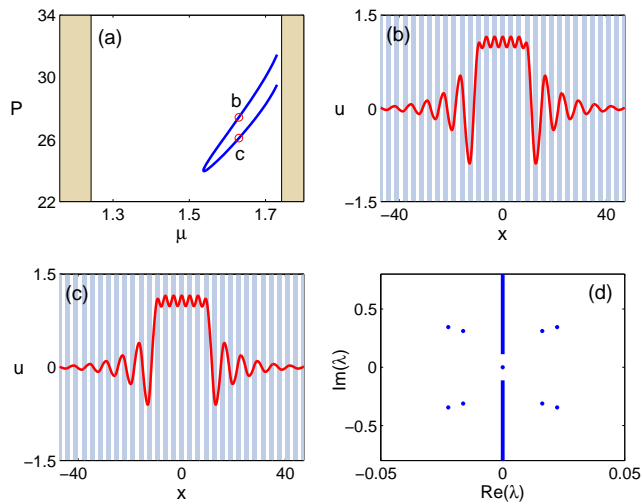}
\end{center}
\caption{Truncated-Bloch-wave solitons in a shallow lattice
potential ($V_0=1$): (a) the power diagram; (b, c) solitons on the
upper and lower branches as marked in (a); (d) linear-stability
spectrum for the lower-branch soliton in (c).} \label{fig5}
\end{figure}

By comparing truncated-Bloch-wave solitons in the deep and shallow
potentials of $V_0=6$ and $V_0=1$, we see that when the potential
becomes shallow, the existence region of these solitons shrinks
toward the upper part of the (first) bandgap (see Figs.~\ref{fig2}
and \ref{fig5}). This shrinkage continues until the potential depth
drops below a critical value of $V_0=0.65$, where these
truncated-Bloch-wave solitons all disappear (see also
Ref.~\cite{our_prl}). Thus very shallow lattice potentials can not
support truncated-Bloch-wave solitons. Regarding the linear
stability of these solitons, the results above show that in deep
lattice potentials, lower branches of truncated-Bloch-wave solitons
are stable, but in shallow potentials, both branches of these
solitons are unstable. As the potential depth decreases, the stable
region of truncated-Bloch-wave solitons gradually shrinks toward the
middle part of the lower branch, and then vanishes when $V_0\approx
2.3$. Thus whole families of these solitons become unstable when
$V_0$ drops below approximately 2.3.  The experiment of
Ref.~\cite{BEC_gap}, which used $V_0 \sim 11$, was thus far above
this instability threshold.  As $V_0$ is defined in the unit of the
lattice recoil energy $E_{rec}$, the dimensionless instability
threshold of $V_0\approx 2.3$ thus corresponds to the dimensional
instability threshold of $V_0 \approx 11.35\hbar^2/md^2$, which is
inversely proportional to the atom mass and the square of the
lattice period.

In this paper, we only analyzed truncated-Bloch-wave solitons in the
first bandgap under repulsive nonlinearity. These solitons were
obtained by truncating the nonlinear Bloch waves originating from
the lower band edge of the first Bloch band. It should be noted that
this branch of nonlinear Bloch waves (see Fig. \ref{fig0}) continues
into higher Bloch bands and bandgaps. In higher bandgaps, by
truncating these nonlinear Bloch waves, we may obtain higher-gap
truncated-Bloch-wave solitons. It is also noted that from every edge
of a Bloch band, a branch of nonlinear Bloch waves bifurcates out
for both attractive and repulsive nonlinearities [the bifurcation
goes rightward (see Fig. \ref{fig0}(a)) under repulsive nonlinearity
and leftward under attractive nonlinearity]. Whenever a branch of
nonlinear Bloch waves enters a bandgap, we may obtain the
corresponding truncated-Bloch-wave solitons. These other
truncated-Bloch-wave solitons can be similarly analyzed, but this
lies outside the scope of the present paper.

In conclusion, we have analyzed the existence and stability of
truncated-Bloch-wave solitons existing in the gaps of the linear
band-gap spectra of matter waves. We have demonstrated that such
self-trapped structures form families of broad but localized
solutions which can be identified by the number of peaks. These
localized states do not bifurcate from the band edge, and their
power diagrams exhibit double branches. Linear stability analysis
demonstrates that in a deep lattice potential, the solutions on the
lower branches are stable regardless of the number of peaks; but in
a shallow potential, solutions on both branches are unstable. When
the lattice potential becomes weak enough, these
truncated-Bloch-wave solitons cease to exist.

This work has been supported by the Air Force Office of Scientific
Research (U.S.A.) and Australian Research Council. The authors thank
Anton Desyatnikov for useful discussions and suggestions. J.Y. also
thanks the Nonlinear Physics Center of the Australian National
University for hospitality during his visit when a part of this work
has been done. His email address is: jyang@cems.uvm.edu.

\end{sloppy}

\begin{thebibliography}{99}
\vspace{-1mm}

\bibitem{optics} D.N. Christodoulides, F. Lederer, and Y. Silberberg, Nature {\bf 424}, 817 (2003).

\bibitem{BEC_new}
A. Trombettoni and A. Smerzi,
%``Discrete Solitons and Breathers with Dilute Bose-Einstein Condensates",
Phys. Rev.  Lett. 86, 2353 (2001).

\bibitem{BEC1}
B. Eiermann, Th. Anker, M. Albiez, M. Taglieber, P. Treutlein, K.P.
Marzlin, and M.K. Oberthaler,
%``Bright Bose-Einstein Gap Solitons of Atoms with Repulsive Interaction",
Phys. Rev. Lett. 92, 230401 (2004).

\bibitem{BEC3}
J.K. Chin, D.E. Miller, Y. Liu, C. Stan, W. Setiawan, C. Sanner, K.
Xu, W. Ketterle,
%``Evidence for Superfluidity of Ultracold Fermions in an Optical Lattice",
Nature 443 , 961-964 (2006).

\bibitem{BEC4}
O. Morsch and M. K. Oberthaler,
%``Dynamics of Bose-Einstein condensates in optical lattices",
Rev. Mod. Phys. 78, 179 (2006).


\bibitem{book} Yu.S. Kivshar and G.P. Agrawal, {\em Optical Solitons: >From Fibers to Photonic
Crystals} (Academic, San Diego, CA 2003), pp. 540.

\bibitem{gap_theory} D.E. Pelinovsky, A.A. Sukhorukov, and Yu.S. Kivshar, Phys. Rev. E {\bf 70}, 036618 (2004).

\bibitem{BEC_gap} T. Anker, M. Albiez, R. Gati, S. Hunsmann, B. Eiermann, A. Trombettoni,
and M.K. Oberthaler, Phys. Rev. Lett. {\bf 94}, 020403 (2005).

\bibitem{our_prl} T.J. Alexander, E.A. Ostrovskaya, and Yu.S. Kivshar, Phys. Rev. Lett. {\bf 96}, 040401 (2006).

\bibitem{BEC_model} F. Dalfovo, S. Giorgini, L. P. Pitaevskii, and S. Stringari, Rev. Mod. Phys. {\bf 71},
463 (1999).

\bibitem{ShiYang} Z. Shi and J. Yang, Phys. Rev. E {\bf 75}, 056602 (2007).

\bibitem{SOM} J. Yang and T.I. Lakoba, Stud. Appl. Math. {\bf 118}, 153 (2007).

\bibitem{torner} E. Smirnov, C.E. Ruter, D. Kip, Y.V. Kartashov, and L. Torner,
%"Observation of higher-order solitons in defocusing waveguide arrays,"
Opt. Lett. 32, 1950-1952 (2007)

\bibitem{jianke_pra} J. Wang and J. Yang, Phys. Rev A {\bf 77}, 033834 (2008).

\bibitem{Boyd} J.P. Boyd, \emph{Chebyshev and Fourier Spectral Methods}, second ed., Dover Publications, 2001.

\bibitem{Yangdipole} J. Yang, I. Makasyuk, A. Bezryadina, and Z. Chen,
Stud. Appl. Math. {\bf 113}, 389 (2004).

\bibitem{panos} P.G. Kevrekidis, H. Susanto and Z. Chen, Phys. Rev. E {\bf 74}, 066606 (2006).

\end{thebibliography}
\end{document}